\documentclass[
    ,final            
  ]
  {aipproc}

\layoutstyle{6x9}

\begin{document}

\begin{flushright}
{\small
hep-ex/0308028\\ 
Fermilab-Conf-03/230-E}\\
\end{flushright}

\title{Future Accelerators(?)
\footnote{Presented at the Conference on the Intersections of Particle
and Nuclear Physics (CIPANP 2003), New York City, May 2003.}  
}

\author{John Womersley\footnote{\tt womersley@fnal.gov}}{
address={Fermi National Accelerator Laboratory, Batavia, IL 60510}
}

\begin{abstract}
I describe the future accelerator facilities 
that are currently foreseen for electroweak scale physics, neutrino
physics, and nuclear structure. I will explore the physics justification 
for these machines, and suggest how the case for future accelerators can be made. 
\end{abstract}

\maketitle


In asking me to give one of the closing presentations at this meeting,
I imagine the conference organizers may expect me to impart some inspiration
as well as information.  The inspirational part will explore 
why I have added a question mark to the title.  The informational
part will describe future accelerators aimed at understanding
electroweak symmetry breaking (TeV scale physics), neutrino 
physics, and nuclear physics.

\section{Why the ?}

The ``?'' indicates that the existence of future accelerators is
far from assured.  In fact, the climate is arguably rather hostile.
In recent years we seem to have done a poor job of making the
case for future machines, at least where particle physics is concerned.
Here are two examples of statements from representatives of the
administration that show how far the case is from being made:

\begin{itemize}
\item
Michael Holland of the White House Office of Management and Budget,
at Snowmass 2001: 
{\it ''How much importance do scientists outside your 
immediate community attach to yoru fervent quest for the Higgs boson? 
How else would you expect us to evaluate your priorities?  What would
you do if the government refused to fund any big accelerator?''}
\cite{holland}

\item
Dr. John Marburger, Director of the Office of Science and Technology
Policy, at SLAC, October 2002:
{\it ``At some point we will simply have to stop building accelerators.  
[...] we must start 
thinking about what fundamental physics will be like when it happens.  
[...] experimental physics 
at the frontier will no longer be able to produce direct excitations 
of increasingly massive parts of nature's spectrum [...]
There are two alternatives.  The first is 
to use the existing accelerators to measure parameters of the 
standard model with ever-increasing accuracy so as to capture 
the indirect effects of higher energy features of the theory [...]
The second is to turn to the laboratory of the cosmos, as physics 
did in the cosmic ray era before accelerators became
available more than fifty years ago.''}
\cite{marb}
\end{itemize}

With all due respect, I have to assert that Dr.~Marburger is wrong on both 
counts.  At some point, yes, any given accelerator technology becomes too 
expensive to pursue.  That does not mean that we have to stop 
building accelerators; it means that we need to develop new
accelerator technologies.  Secondly, the very richness of the
``laboratory of the cosmos'' is exactly the reason why we need 
to keep building accelerators.  There is a universe full of
weird stuff out there --- the more we look, the more weird stuff
we find.  Do we really think we can understand it all without making
these new quanta in the laboratory and studying their properties
under controlled conditions?

How might we then start to better make the case?  I have a couple of
suggestions.

\subsection{1. Emphasize the Unknown}

As Shakespeare had Hamlet point out, ``there are more things in Heaven 
and Earth than are dreamt of in our philosophy.''  In justifying
and describing the potential of new facilities, I believe that
we have tended much too far in the direction of `one last piece
of the puzzle' or `we know what
we're doing and we know what we'll find.' This reinforces the
mistaken
idea that we are close to `the end of science' and is rather
hard to justify given that 95\% of the universe is not made
of quarks and leptons.  In fact, exploring the unknown has
a lot more resonance with the public.  We have to search for
new phenomena in ways that are not constrained by our
preconceptions of what may be `out there.'  The Tevatron
collider experiments have done just that. The D\O\ collaboration
has published \cite{sleuth} a model-independent search for
deviations from the standard model in the 1992-95 data.
Only two channels had any hint of disagreement and overall
the confidence level for the standard model---in this small
dataset---was 89\%.  
CDF has also pursued signature-based searches.  
Such approaches are good science
but also good tools for publicity and outreach.

\subsection{2. It's all about the Cosmos}

The composition of the universe is a powerful unifying theme 
for particle and nuclear physics.  Mass shapes the universe
through gravity, the only force that is important over astronomical
distances.  The masses of stars and planets arise largely through
QCD (binding energies of protons and neutrons), but it has long
been known that there is substantial invisible (dark) matter
and that (from primordial D/He abundances) that this matter is
not baryons.  Recent measurements of the multipole moments
of the cosmic microwave background
such as that from WMAP\cite{wmap} have allowed the dark matter 
density to be extracted quite precisely.  There seems to be
about six to seven times more mass ($27\pm 4$\%) than
baryons ($4.4\pm 0.4$\%).  The most likely explanation is
that the dark matter is a new kind of particle: weakly
interacting, massive relics from the early universe.  There
are two complementary experimental approaches that should 
be pursued: to search for dark matter particles impinging on
Earth, and to try to create such particles in our accelerators.

Supersymmetry (SUSY) is an attractive idea theoretically; it can 
unify couplings, cancel divergences in the Higgs mass, and
provides a path to the incorporation of gravity and string
theory. It also predicts a particle, the lightest neutralino,
which is a good explanation for
cosmic dark matter and which could be discovered at the
Tevatron or LHC, and studied in detail at a linear collider.
In fact the search for dark matter is underway now, in Run~II
at the Tevatron collider.  Neutralinos would be produced in
cascade decays of squarks or gluinos and could be
detected through their escape from the detector, as 
missing transverse energy. 

The same cosmic microwave background data, together with 
supernova measurements of the velocity of distant galaxies,
suggest that two-thirds of the energy density of the universe 
is in the form of dark energy---some kind of field that 
expands along with the universe.  Again, there are two
complementary approaches to learn more.  We should refine
our cosmologically-based understanding of the properties
of dark energy in bulk (its `equation of state') through
new projects such as SNAP.  We should also understand what
we can do under controlled conditions in the laboratory.
Ultimately I am sure we will want to make dark energy
quanta in accelerators. 
For now, we should explore the only other
example of a `mysterious field that fills the universe,'
namely the Higgs field.  The Standard Model Higgs field would 
produce something like 54 orders of magnitude too much dark 
energy compared with the cosmological observations, but surely
it cannot be totally unrelated.  

We know that photons and $W$ and $Z$ bosons couple to
particles with the same strength---this is electroweak unification. 
Yet while the whole universe
is filled with photons, the $W$'s and $Z$'s only mediate a weak
force that occurs inside nuclei in radioactive beta decay.  
This is because the $W$ and $Z$ are massive particles, and
the unification is thus broken.  This mass (the electroweak
symmetry breaking) appears to arise because the universe is
filled with an energy field, called the Higgs field, with 
which the $W$ and $Z$ interact (and in fact mix).  We want to
excite the quanta of this field and measure their properties.  
The field need not result from a single, elementary scalar
boson: there can be more than one particle (as is the case
in supersymmetry), or composite particles can play the role of
the Higgs (e.g. in technicolor or topcolor models).  We do
know that electroweak symmetry breaking occurs, so there is
something out there coupling to the $W$ and $Z$.  Precision
electroweak measurements imply that this thing looks very much like
a standard model Higgs (though its couplings to fermions are
less constrained).  We also know that $WW$ cross sections
would violate unitarity at $\sim 1$~TeV without it, and this
is a real process that will be seen at the LHC.  For all of 
these reasons, electroweak symmetry breaking remains a
focus of the experimental high energy physics program.

This naturally leads me to the second part of my 
presentation, where I
shall review future accelerator intitiatives, starting 
with those aimed at the electroweak scale.

\section{Future Accelerators for 
Electroweak Scale Physics}

The flagship future facility for TeV-scale physics will be the
{\bf Large Hadron Collider} (LHC) at CERN.  The LHC is a 14~TeV
proton-proton collider.  It will serve two large general 
purpose detectors, ATLAS and CMS, together with a heavy-ion 
and $B$-physics program.
Underground construction is well advanced and the detectors
are making good progress.  Accelerator dipole magnet production 
is the overall pacing item; if all goes well, first beam will be
circulated in 2007.

The LHC will be able to discover a standard model Higgs over
the entire range of allowed masses (115~GeV -- 1~TeV).  
Beyond discovery, we will need to verify that the observed state
actually provides both vector bosons and fermions with
their masses.  The LHC will be able to start this job
by measuring various ratios of Higgs couplings and
branching fractions (at the 25\% level)
by comparing rates in different Higgs production and decay
channels.

The more complex Higgs sector in supersymmetric models
can also be quite thoroughly explored.  Tau decay modes
are very important over a large region of parameter space
at moderate to large $\tan \beta$.
At least one Higgs state is visible no matter what; the
most problematic region of parameter space is where one
light state $h$ is discoverable, but looks very much
like the standard model $H$. 

To elucidate this case, and of course in general too, one
would use the LHC to search for supersymmetry through
sparticle production.  The mass range covered for squarks
and gluinos is huge (up to $\sim 2.5$~TeV) and a signal
to background ratio as high as ten can be achieved even with
simple cuts.  Exclusive mass reconstruction of SUSY
cascade decays has been demonstrated for several 
benchmark points.  New Higgs signals also appear in such
decays.

The combination of high energy (14~TeV) and luminosity (100~fb$^{-1}$)
means that the LHC will have the potential to observe almost
any other new physics associated with the TeV scale.  Extra
dimensions of space-time and/or TeV-scale gravity could 
have subtle, indirect effects---or direct, spectacular signatures
like the production of black holes.  The LHC would also
be sensitive to compositeness, excited quarks, leptoquarks,
technicolor, strong $WW$ interactions, new gauge bosons, and 
heavy neutrinos.

In summary, by the year $201x$, if all goes well, we should
have observed at least one and maybe several Higgs bosons, and
will have tested their properties at the 25\% level.  We will
not always have been able to distinguish a Standard Model from
a SUSY Higgs, but we almost always expect to have discovered 
SUSY in other ways.  If we don't see a Higgs, we will have
observed some other signal of electroweak symmetry breaking
(technicolor, or strong $WW$ scattering, for example).  In
addition, we will have learned a great deal more about the 
physics landscape at the TeV scale: is there supersymmetry?
Are there extra dimensions?

There is an international consensus\cite{hepap} that the highest priority
facility to follow the LHC should be an 
{\bf Electron-positron Linear Collider} (LC).  
This would
collide $e^+e^-$ beams at a center-of-mass energy 
between 500~GeV and 1~TeV and deliver a few hundred inverse
femtobarns per year.  The cost is perhaps {\$}5--7B and will
require an international effort to build; it could be in
operation by 2015--20.

The physics of the Linear Collider is no longer about discovery,
it is about precision.  (In this sense, it plays a similar role
to the one that LEP did, after the $W$ and $Z$ had been
discovered at the SPS collider).  The LC program aims to
exploit aggressive detector technology such as displaced
vertex charm-tagging and energy-flow calorimetry, and also make
use of highly polarized beams to reduce backgrounds.

Higgs production at a LC occurs through both
$e^+e^- \rightarrow HZ$ and $e^+e^- \rightarrow \nu\overline\nu H$
processes.  The $HZ$ process can be used to reconstruct the 
Higgs (actually whatever the $Z$ recoils against) even if
it decays invisibly, and permits the $g_{HZZ}$ coupling to be
determined to a few percent.  This in turn provides a simple
test of whether the observed particle is actually the only Higgs:
namely, does it account for all the mass of the $Z$?  For
example, in minimal SUSY the 
$h$ couples $g_{hZZ}\sim g_Z M_Z \sin(\beta - \alpha)$
and the $H$ couples $g_{HZZ}\sim g_Z M_Z \cos(\beta - \alpha)$ and 
together they create the full $M_Z$ that we observe.
The $\nu\overline\nu H$ process, with $H\to b\overline b$,
allows the $g_{HWW}$ coupling to be extracted with a precision of
a few percent.  

The couplings of the Higgs to fermions determine whether the
Higgs field is indeed responsible for fermion masses as well as
for electroweak symmetry breaking.  With 500~fb$^-1$ at
$\sqrt{s}=500$~GeV, the Yukawa couplings of a 120~GeV Higgs could 
be determined at the level of 
$\Delta g_{Hbb} = 4$\%, 
$\Delta g_{Hcc} = 7$\%, 
$\Delta g_{H\tau \tau} = 7$\%, and 
$\Delta g_{H\mu \mu} = 30$\%.  
At $\sqrt{s}=800$~GeV, it would also be possible to
measure $g_{Htt}$, through $t\overline t H$ production, 
at the 10\% level. We could thus determine
whether the top quark's unexpectedly large mass arises
from the Higgs or from some other mechanism.

The quantum numbers of the Higgs itself can be excplored.
The angular dependence of $e^+e^- \to ZH$ and of the 
$Z \to f\overline f$ decay products can cleanly separate
$CP$-even and odd Higgs states ($H$ and $A$ in minimal
supersymmetry).  One would be sensitive to a 3\% admixture
of $CP$-odd $A$ in the ``$H$'' signal.  This could be a window
to $CP$ violation in the Higgs sector.  
With sufficient luminosity, the Higgs self-coupling can be 
probed through $ZHH$ production (six jets in the final state).
The cross section is tiny, about 0.2~fb, so of order 1~ab
(1000~fb) is required for a 20-30\% measurement of $g_{HHH}$.
Such a measurement would constrain the Higgs potential and,
compared with the expectation from the Higgs mass, would
give a self-consistency test for the Higgs.

There are very clean signals for light superpartner production 
at a LC.  For example, chargino pair production occurs through
$s$-channel annihilation or through $t$-channel sneutrino
exchange.  One can select the mixture of processes by polarizing
the electron beam: since a right-handed electron has no coupling
to the sneutrino, one suppresses the $t$-channel process. In this
way the ``Wino'' and ``Higgsino'' parts of the chargino can
be separated. The Wino coupling to $e\tilde \nu$ can then be
compared to the $W$ coupling to $e\nu$ --- if it is truly
supersymmetry, they must be equal.  The chargino decays to
neutralinos, and at the LC all the masses can be measured.
This would enable the expected dark matter abundance and
properties to be calculated.

In summary, we are planning a relay race at the electroweak scale.
The Tevatron will discover new TeV-scale physics if we are 
lucky.  The LHC is ``guaranteed'' discovery and will start to
measure and constrain.  The Linear Collider will measure,
measure, measure --- and build the physics case for the next
accelerator to follow.

\section{Future Accelerators for Neutrino Physics}

We now have three distinct signals for neutrino oscillation:
\begin{itemize}
\item {\bf Solar neutrinos:} missing $\nu_e$, as observed by
Homestake, GALLEX, SAGE, Kamiokande, SuperK, SNO and KamLAND.
\item {\bf Atmospheric neutrinos:} missing $\nu_\mu$, as
observed by Kamiokande, SuperK and K2K.
\item {\bf LSND signal:} a $\nu_\mu \leftrightarrow \nu_e$ oscillation, 
as seen by the LSND experiment at Los Alamos.
\end{itemize}
Parenthetically, we may note (and point out to Dr. Marburger) that
while the ``laboratory of the solar system'' gave us the
first two signals, it required terrestrial beams (at KamLAND and
K2K) to really understand and have confidence in what we were seeing.

The solar and atmospheric signals form a consistent picture
in which three neutrino mass eigenstates each contain admixtures
of the flavor states.  The $\nu_1$ and $\nu_2$ states are separated by 
$\Delta m^2 \sim 5\times 10^{-5}$~eV (the solar oscillation
signal) while $\nu_3$ is split from these two states by 
$\Delta m^2 \sim 3\times 10^{-3}$~eV (the atmospheric oscillation
signal).  The overall mass scale and ordering in mass is
not known.  Unlike quarks, there is a lot of mixing; 
the mass eigenstates do not correspond ``mostly'' to any single 
flavor.  If the LSND result is confirmed, it would require
drastic extensions to this picture: either additional neutrino
states, or new physics ($CPT$ violation, for example).

There are a significant number of neutrino experiments now
running. At Fermilab, miniBooNE is seeking to confirm LSND's
signal for $\nu_\mu \to \nu_e$ (and 
also $\overline \nu_\mu \to \overline \nu_e$).  In Japan, K2K
is pursuing the ``atmospheric'' oscillation using an accelerator
neutrino beam, and KamLAND is exploring the ``solar'' signal
using reactor neutrinos.  SNO continues to detect solar
neutrinos with flavor selection.  It will soon
be joined by Borexino, a solar neutrino detector with a very
low energy threshold.  Two new long-baseline projects are
also under construction: MINOS, with a beam from Fermilab to
Soudan to measure the atmospheric oscillation and search for
$\nu_\mu \to \nu_e$; and the CERN Neutrinos to Gran Sasso project
which will focus on  $\nu_\mu \to \nu_\tau$ using the OPERA
(emulsion)
and ICANOE (liquid argon) detectors.

These experiments will tell us whether the LSND result is
correct (if yes, confirming that there is new physics).
They will better pin down the mass-squared splittings
and mixing angles in the solar and atmospheric oscillations.
Most importantly, they will give some information on the
critical parameter $\theta_{13}$, which describes how much
electron-neutrino there is in the $\nu_3$ eigenstate.
It is  $\theta_{13}$ which governs the size of possible
$CP$ violation in the neutrino sector, which is of great
interest in understanding the baryon asymmetry of the
universe.  Currently,  $\theta_{13}$ is known to be less
than about 0.10.  If it is large enough (where large
enough means greater than 0.05 or so), a rich program
of next generation experiments opens up.  The goal would
be to search for electron neutrinos in the ``atmospheric'' 
distance/energy regime, to observe matter effects
(to resolve the mass ordering) and ultimately $CP$ violation.
This would require any or all
of the following:
\begin{itemize}
\item Bigger detectors, 20-100~kt compared with MINOS's 3~kt
fiducial mass;
\item Better instrumentation (for example, calorimetry);
\item Higher intensity neutrino beams (``superbeams'').
\end{itemize}
There are a number of concepts that exploit new beams
to existing detectors, or new detectors in existing beams,
or entirely new projects:  Fermilab to Minnesota or Canada,
Brookhaven to Homestake or WIPP, and JPARC to Kamioka.
One could also access the physics through $\nu_e$ 
disappearance using a very high precision reactor experiment.  

If $\theta_{13}$ is small, things become much more challenging.
Baselines of thousands of kilometers become optimal, and low rates
require new technology for neutrino beams.  In this scenario, 
a muon storage ring neutrino factory may be essential.

No matter what we learn in the next few years, it is clear
that we will need major new accelerator and detector facilities
for neutrino physics.  There is no complete 
consensus---yet---on just what those facilities should be, but
there are lots of good ideas, and lots more data are coming.

\section{Future Accelerators for Nuclear Physics}

The nuclear physics community has developed a long range plan
for the next decade\cite{nucplan}, and recently the Facilities
Subcommittee of the Nuclear Science Advisory Committee reported
on the importance of the science and readiness for construction
of new facilities\cite{nucfac}.  
The following three projects were the highest 
ranked in the two categories:
\begin{itemize}
\item The Rare Isotope Accelerator (RIA)
\item A new gamma-ray detector array GRETA (instrumentation for RIA)
\item CEBAF energy upgrade (from 6 to 12~GeV).
\end{itemize}
(RHIC upgrades and an underground detector were also highly ranked
but not judged to be immediately ready for construction.)

RIA is a facility to produce rare isotopes.  It is driven
by a linac (400~MeV$/u$ U, 900~MeV $p$) which feeds production
targets followed by online isotope separation, possible
re-acceleration, trapping or isotope recovery.  So why do we need 
such a major ($\sim$\$900M) new facility for nuclear physics now?  
The science case is based on:
\begin{itemize}
\item Nuclear struture;
\item Astrophysics --- the origin of elements heavier than iron. 
Creation of such elements in supernovae is believed to occur 
through a complex series of reactions involving unstable,
neutron-rich nuclei that could be explored at RIA;
\item Low energy tests of standard model symmetries.
\end{itemize}
As well as the science, RIA would offer ``collateral benefits''
through the production of medical isotopes and the understanding
of processes relevant to nuclear stockpile stewardship. 

In preparing this talk I discussed the RIA science case 
with several of my high energy physics
colleagues.  Their initial skepticism generally turned to interest
once they heard the astrophysics aspects (and, implicitly,
how much had been glossed over in the undergraduate astronomy 
classes they had taken).
This observed resonance is a good lesson for all of us in how
to explain the relevance and interest of future facilities to
those outside our immediate field.

\section{conclusions}

Accelerators are the key to understanding this weird and wonderful
universe that we inhabit.  Only accelerators can provide the controlled
conditions, known particle species, high rates and high energies 
that we need to make sense of cosmological observations.
Recent progress in astroparticle physics and cosmology does
not weaken the case for new accelerators, it strengthens it;
and there is no shame in exploiting public interest in these
discoveries.  The major problems are political.  As Joe Lykken
stated at the Lepton-Photon Symposium at Stanford in 1999, 
``It is much more likely that we will fail to build new
accelerators than that these accelerators will fail to
find interesting physics.''
It will take a concerted effort to overcome the political
obstacles, but if we work together we can do it.

\begin{theacknowledgments}
I would like to thank Peter Meyers for allowing me to use 
material from his excellent presentation at the 2003 meeting
of the American Physical Society in Philadelphia.
\end{theacknowledgments}

\end{document}